# Two Regime Cooling in Flow Induced by a Spark Discharge

Bhavini Singh[1], Lalit K. Rajendran[1], Pavlos P. Vlachos[2] and Sally P.M. Bane[1]

**Abstract**

The cooling process associated with the flow induced by a spark plasma discharge generated between a pair of electrodes is measured using stereoscopic particle image velocimetry (S-PIV) and background oriented schlieren (BOS). Density measurements show that the hot gas kernel initially cools fast by convective cooling, followed by a slower cooling process. The cooling rates during the fast regime range from being 2 to 10 times those in the slower regime. An analytical model is developed to relate the cooling observed in the fast regime from BOS, to the total entrainment of cold ambient fluid per unit volume of the hot gas kernel, measured from S-PIV. The model calculates the cooling ratio to characterize the cooling process and shows that the cooling ratio estimated from the density measurements are in close agreement with those calculated from the entrainment. These measurements represent the first ever quantitative density and velocity measurements of the flow induced by a spark discharge and reveal the role of entrainment on the cooling of the hot gas kernel. These results underscore that convective cooling of the hot gas kernel, in the fast regime, leads to approximately 50% of the cooling and occurs within the first millisecond of the induced flow.

1. ## Introduction

Plasma discharges are created by the breakdown of a gas in the presence of a strong electromagnetic field, leading to the formation of ionized species which are on average electrically neutral. These discharges have widespread applications in flow and combustion control, material processing, biomedicine, nanotechnology, and environmental engineering [1]–[5]. In flow control applications, the discharges are typically used to induce localized heating and fluid motion to control and modify transport processes. The induced flow can be produced through an electrohydrodynamic mechanism such as in dielectric barrier discharge (DBD) actuators where a body force is generated due to the electric field thereby inducing a wall jet [3], [6]–[8], or via a thermal mechanism such as in spark discharge actuators.

Spark discharges are typically generated by raising the voltage difference between two electrodes, until the breakdown voltage is reached, resulting in ionization of gas in the electrode gap. The spark initiates a kernel of gas at very high temperature and pressure which then expands outward [9], producing a shock wave that lasts on the order of a few microseconds, which then gives rise

[1] School of Aeronautics and Astronautics, Purdue University, West Lafayette, IN 47907, USA
[2] School of Mechanical Engineering, Purdue University, West Lafayette, IN 47907, USA

to a transient three-dimensional flow field [10]. The flow induced by spark discharges have been the focus of a few computational studies [11]–[13] as well as some experimental investigations, e.g., [14], [15]. Computations on the flow induced by a single spark have thus far been restricted to the very early stages (< 100 µs) of flow development [16]. While these studies showed the presence of a torus-shaped hot gas kernel and a weak shock, the development of the kernel in the post shock phase of the flow, and the associated heat, mass and momentum transport mechanisms have not been characterized. Experimental results of the late stages of the flow induced by sparks have been restricted to qualitative descriptions of the flow field [15], primarily due to difficulties associated with measuring such a transient and highly complex flow field with sufficient spatiotemporal resolution. Some of the first experimental results presenting quantitative information on the flow induced by microsecond-duration sparks [17] have shown the presence of a pair of vortex rings around the hot gas kernel for 5 mm and 8 mm electrode gap distances. In addition, these experiments revealed, jets of surrounding gas entrained into the electrode gap, and the circulation of the vortex rings as well as the rate of entrainment was found to decay over time. The rates of decay were similar for both electrode gaps, when non-dimensionalized using the electrode gap distance and velocity induced behind the shock produced by the spark.

While spark plasma discharges are being applied for various flow control and combustion applications [18]–[20], a detailed characterization of the flow induced by these discharges under quiescent conditions is still lacking. For example, recent work on the applications of NRP discharges to ignition have shown that the pulse frequency and number of pulses have a distinct effect on the size of the flame kernel, and this effect was attributed to a possible coupling between the characteristic recirculation time from the induced flow and the inter-pulse time interval [21]. Detailed quantitative characterization of the flow induced by these discharges under quiescent conditions is a necessary step in characterizing these actuators to enable informed actuator designs, tailored to specific applications.

The purpose of this paper is to explain the cooling process observed in spark plasma discharge induced flow fields. This will be done by characterizing the flow induced by a single nanosecond-scale spark discharge using high speed stereo Particle Image Velocimetry (S-PIV) and Background Oriented Schlieren (BOS) to obtain velocity and density measurements in the post-shock phase of the induced flow. Using density information from the BOS measurements, it will be shown that there are two distinct cooling regimes of the spark-induced hot gas kernel with significantly different cooling rates. Further, the cold ambient gas entrained into the electrode gap leads to a bulk of the cooling of the hot gas kernel in the first regime. To assess this effect, a model is developed to relate the cooling rate of the hot gas kernel to the volume of the entrained fluid. The measurements show good agreement with the model within the limits of experimental uncertainty.

## 2. Experimental methods and techniques

To investigate the flow induced by a spark plasma, two separate but complementary experiments were conducted. In the first test, velocity measurements of the flow induced by a nanosecond spark discharge were obtained using stereoscopic particle image velocimetry (S-PIV) measurements, and in the second the density of the hot gas kernel was measured under the same spark generating conditions, using background oriented schlieren (BOS). In each test, 25 runs were conducted, i.e.

25 separate spark plasma discharges were measured. Each run was spaced out in time such that there were no residual flow effects from one spark event to the next, that is, each run was at least 30 seconds apart. Voltage and current measurements were taken with each run.

### 2.1  Plasma generation – pulser and electrode description

A nanosecond high voltage pulse generator from Eagle Harbor Technologies was used to generate a spark discharge between two electrodes. The pulse parameters of the pulse generator can be independently varied, with peak voltages up to 25 kV, pulse durations from 20 to 110 ns, and pulse repetition frequencies (PRF) up to 400 kHz. The electrodes, shown in Figure 1, were machined out of ceriated-tungsten and had cone shaped tips with radius of curvature of approximately 150 µm to enhance the electric field and aid in breakdown. The electrodes were separated by 5 mm in atmospheric, quiescent air. The voltage and current across the electrode gap during breakdown and spark discharge were measured using two Tektronix P6015A high voltage probes (in a differential measurement configuration) and a Magnelab CT-D1.0 current transformer, respectively. These measurements were used to calculate the electrical energy deposited in the plasma as shown in Figure 1 (c). The electrical energy deposited in the plasma ranged from 4 mJ to 7 mJ.

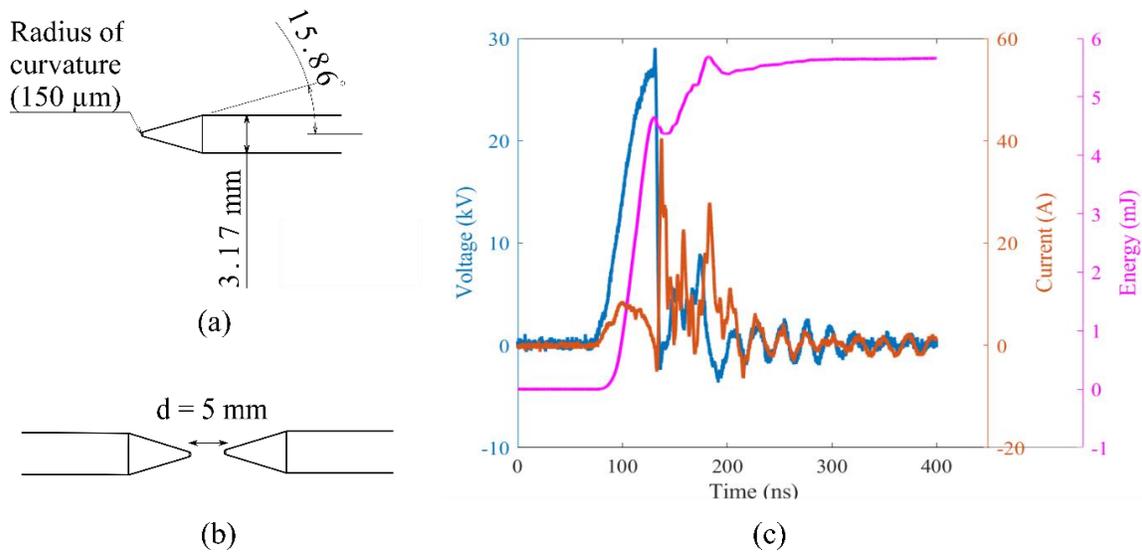

Figure 1: (a) Sketch of the cone-tipped electrode, (b) the two-electrode configuration used for plasma generation, and (c) sample current, voltage and energy waveform produced during a spark discharge in a 5 mm electrode gap.

### 2.2  Stereoscopic Particle Image Velocimetry
#### 2.2.1  Experimental Setup

A schematic of the stereoscopic particle image velocimetry (S-PIV) system used to analyze the flow field is shown in Figure 2. The main components of the set-up include a Q-switched EdgeWave Nd: YAG laser which was operated at 10 kHz with pulse separation of 30 µs, two high speed cameras (Photron SA-Z camera and Phantom v2512), and an acrylic test section measuring 190.5 x 140 x 152 mm, containing the electrodes. The laser sheet optics produced an approximately 1 mm thin waist in the region of interest where the spark discharge was generated. A Quantum Composer Model 575 delay generator and LaVision high speed controller were used to synchronize and trigger the laser, cameras, and high voltage pulse generator. The Photron and Phantom cameras with Nikon Nikorr 105 mm lenses were used to record particle images at 20,000 fps (corresponding to 20 kHz sampling rate). The resolutions for the two cameras were 1024 x 1024 pixels for the Photron and 1280 x 800 pixels for the Phantom, with approximate fields of view of 10 x 10 mm and 20 x 13 mm, respectively. The included angle between the two cameras was $40^0$. A fluidized bed seeder was used to inject aluminum oxide particles with diameters of about 0.3 µm and estimated Stokes number of approximately 0.002 into the test section.

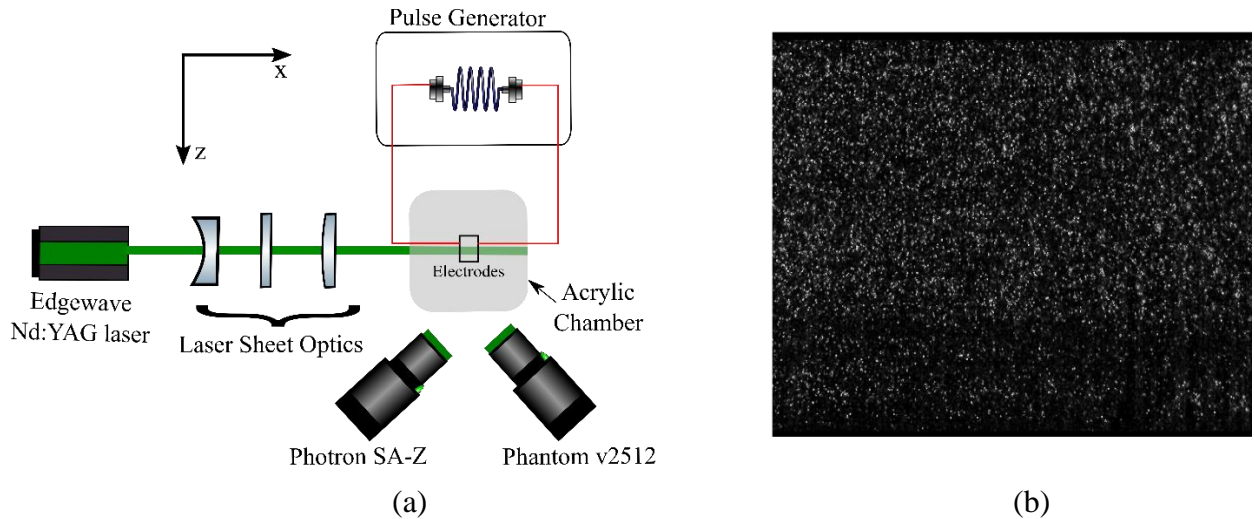

Figure 2: (a) Schematic of the experimental set-up for S-PIV measurements of the plasma-induced flow field, (b) recorded S-PIV image.

### 2.2.2 Image Processing

PRANA (PIV, Research and ANAlysis) software (https://github.com/aether-lab/prana) developed by Vlachos and coworkers at Virginia Tech and Purdue University was used to process the recorded particle images. For the S-PIV measurements, a calibration was first performed to calculate the mapping function of the camera. The camera coordinates and physical coordinates were fitted using a least squares polynomial with cubic dependence in the $x$ and $y$ directions and quadratic dependence in the $z$ direction [22]. A self-calibration procedure [23] was then followed to account for misalignment between the calibration target and the plane of the laser sheet, and the mapping function was corrected accordingly. With the new mapping function and calibration information, the individual camera images from each camera were dewarped onto a common grid and cross correlated to yield particle displacement values [24]. The correlation method used was the Robust Phase Correlation (RPC) [25]–[27] in an iterative multigrid framework using window deformation [28]–[31], with each pass validated by universal outlier detection (UOD) [32]. A total of four passes were used and a 50% Gaussian window was applied to the original window size [26] resulting in window resolutions of 64x64 pixels in the first pass to 32x32 pixels in the last pass, with 50 % window overlap in all passes. Between successive passes, velocity interpolation was performed using bicubic interpolation and the image interpolation was performed using a sinc interpolation with a Blackman filter. Finally, subpixel displacement was estimated using a three-point Gaussian fit [33]. The projected velocity fields calculated in this manner were then combined with the camera angle information obtained from calibration to yield the three components of velocity in the measurement plane [34]. Proper orthogonal decomposition with the entropy line fit (ELF) method [35] was used to denoise the PIV velocity fields. The spatial resolution for velocity calculations was 0.32 mm and each snapshot contains 50 x 64 vectors.

*2.2.3   Uncertainty quantification*

Experimental uncertainties vary in space and time, and are unique to each experiment [36]. Various *a-posteriori* methods have been developed to quantify systematic and random uncertainty bounds in PIV measurements, grouped into indirect methods [37]–[39] and direct methods [40]–[42] of uncertainty quantification. Experimental uncertainties in the S-PIV measurements were propagated according to the procedure outlined by Bhattacharya *et al* [43]. Uncertainties in the planar velocity fields were calculated using Image Matching (IM) [41] which uses the position disparity between particle pairs in the two images at the end of a converged deformation processing to estimate the uncertainty in the displacement value. IM was used as it is a direct method and these have been shown to be more sensitive to elemental error sources [44]. However, the IM method has some drawbacks due to its reliance on pairing of individual particle images which can be imprecise in highly seeded images or can be inaccurate due to out of plane motion [36]. Any of the other direct uncertainty quantification methods could be used though each method has its own drawbacks and planar PIV uncertainty quantification is an active area of research. These uncertainties were then propagated through the stereo reconstruction by combining uncertainties in the planar velocity field with positional uncertainties calculated from the disparity map obtained after the self-calibration procedure and uncertainties in the camera angles, to calculate uncertainty in the reconstructed velocity field. Uncertainties in the three components of velocity (u, v and w) are on average 1.2%, 1% and 3% of the maximum velocity (3 m/s), respectively.

## 2.3   Background Oriented Schlieren
*2.3.1   Experimental Setup*

Background Oriented Schlieren (BOS) was used to measure density gradients in the post-discharge flow by tracking the apparent distortion of a target dot pattern. The dot pattern was generated by illuminating a block of sand-blasted aluminum with a 150 W xenon arc lamp (Newport 66907), and the images were captured at 20 kHz using a Photron SA-Z camera with a 105 mm Nikon lens with an f# of 11 and a 2X tele-converter for an approximate field of view of 14 x 14 mm. A schematic of the BOS system is shown in Figure 3. The distance from the center of the electrodes to the BOS target ($Z_D$) was approximately 57 mm and the distance from the center of the electrodes to the camera ($Z_A$) was approximately 152 mm.

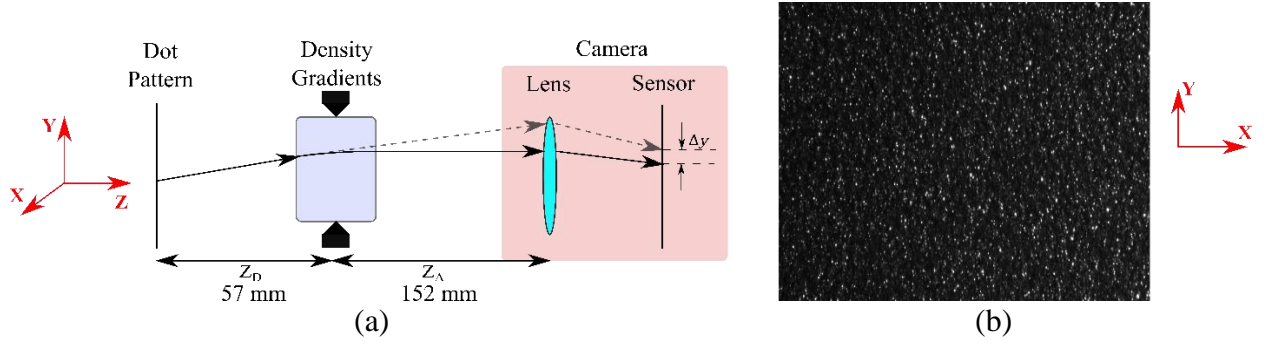

Figure 3: (a) Experimental schematic of the primary components of the BOS system; (b) recorded BOS image

### 2.3.2 *Image Processing*

The distortion of the dot pattern is estimated by cross-correlating an image taken with the flow to a reference image taken without the flow. The images were processed using the same approach and software (PRANA) as for the S-PIV measurements. The correlation method used was the Robust Phase Correlation (RPC) [25]–[27] in an iterative multigrid framework using window deformation [28]–[31]. Each pass was validated by universal outlier detection (UOD) [32]. A total of four passes were used with window resolutions of 64x64 pixels in the first pass to 32x32 pixels in the last pass, with 50 % window overlap in all passes. Subpixel displacement was estimated using a three-point Gaussian fit [33]. The spatial resolution of the final pass was 0.2 mm and each snapshot contains 22 x 55 vectors.

The pixel displacements obtained from the cross-correlation analysis were then used to calculate the projected density gradient field [45],

$$\frac{\partial \rho_p}{\partial x} = \int \frac{\partial \rho}{\partial x} dz = \frac{\Delta x}{Z_D M} \frac{n_0}{K} \tag{1a}$$

$$\frac{\partial \rho_p}{\partial y} = \int \frac{\partial \rho}{\partial y} dz = \frac{\Delta y}{Z_D M} \frac{n_0}{K} \tag{1b}$$

where $\partial \rho / \partial x$ is the density gradient along the $x$ direction, $\rho_p = \int \rho dz$ is the projected density field, $\Delta x$ is the pixel displacement on the camera sensor, $M$ is the magnification, $Z_D$ is the distance between the mid-plane of the density gradient and the dot pattern, $n_0$ is the refractive index of the undisturbed medium, and $K$ is the Gladstone-Dale constant.

The projected density field $\rho_p(x,y)$ was calculated by 2D integration, on solving the Poisson equation given by:

$$\frac{\partial^2 \rho_p}{\partial x^2} + \frac{\partial^2 \rho_p}{\partial y^2} = S(x,y) \tag{2}$$

Dirichlet boundary conditions were used on the left and right boundary, with $\rho_p = 0$ at the edges since there was no flow due to the spark in these regions. The field of view was chosen to be about one electrode gap along the radial direction on each side, such that the left and right boundaries were far enough from the discharge to justify this boundary condition. At the top boundaries, the displacements obtained from the cross-correlation were integrated along the boundaries starting from the left and right corners with a known $\rho_p$ to set up an *artificial* Dirichlet boundary condition. The Poisson equation was solved iteratively until convergence to obtain the final projected density field.

Finally, the actual density field $\rho(r,z)$ was calculated from the projected density field by Abel inversion under the assumption that the flow was approximately axisymmetric [46]. This was a reasonable assumption in this problem because the geometry is axisymmetric. However, given that the spark channel may not be axisymmetric, the post-discharge flow may deviate from perfect axisymmetry, and this is a possible source of error in these measurements.

For the 2D axisymmetric field shown in Figure 4, the Abel inversion equation is given by,

$$\rho(r) = -\frac{1}{\pi} \int_r^\infty \frac{d\rho_p}{dx} \frac{dx}{\sqrt{x^2 - r^2}} \tag{3}$$

and can be used to calculate a 2D axisymmetric field $\rho(r)$ from a 1D projection $\rho_p(x)$. Since the BOS measurements are 2D projections, this procedure was sequentially applied for each horizontal row of measurements to reconstruct 2D slices of the full 3D axisymmetric flow field.

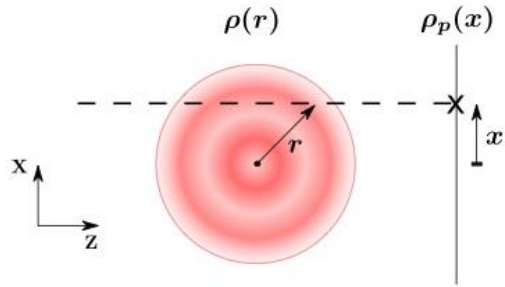

Figure 4: Projection of a 2D axisymmetric field.

### 2.3.3 *Uncertainty quantification*

Experimental uncertainties in the BOS measurements were propagated through the complete density reconstruction according to [47]. Uncertainties in the displacement field were calculated using Image Matching [41], then propagated through the optical layout followed by the Poisson solver to obtain uncertainty in the projected density field. The uncertainties were then propagated through the Abel inversion to obtain total uncertainty in the density field. Uncertainties in the density field were on average 0.01% of ambient density (1.225 kg/m$^3$).

## 2.4 Post processing metrics

The measured velocity and density fields were used to calculate additional quantities and metrics to characterize the induced flow. From the velocity field obtained using S-PIV measurements, the vorticity field was calculated using a 4th order compact noise optimized Richardson extrapolation scheme [48]. The coherent structures in this vorticity field were then identified using swirl strength, using the $\lambda_{ci}$ method [49]. All swirl values that were at least 5% of the maximum swirl strength and had a specified minimum area were considered coherent structures.

From the calculated density field, the kernel is identified as any region of density that is lower than 95% of ambient density (1.225 kg/m³). The mean density at a single snapshot was calculated by summing all the density values in the identified kernel and dividing this by the number of points in the kernel.

## 2.5 Non-dimensionalization

Results were non-dimensionalized similar to prior work [17] using the velocity induced behind the shock wave produced by the spark expansion. The shock was assumed to be weak [50], [51] and the weak shock solution for the pressure jump across a shock, developed by Jones *et al* [52], was used:

$$\frac{p_2 - p_1}{p_1} = \frac{2\gamma}{\gamma + 1} \frac{0.4503}{(1 + 4.803(r/R_0)^2)^{\frac{3}{8}} - 1} \tag{4}$$

where the subscripts 1 and 2 represent conditions upstream and downstream the shock, respectively. The quantity *p* represents pressure, and *r* is the radial distance at which flow properties behind the shock are measured. In the present context, the radial distance was considered to be the electrode gap distance (5 mm). The variable $R_0$ is the characteristic radius determined by the initial conditions, and is given by:

$$R_0 = \left[\frac{4E_0}{3.94\gamma p_1}\right]^{\frac{1}{2}} \tag{5}$$

where $E_0$ is the energy deposited by the plasma per unit length of the electrode gap and $\gamma$ is the ratio of specific heats (taken to be 1.4 for air). Using these equations, the upstream Mach number and the shock speed were determined, and then the normal shock relations were used to calculate the induced velocity and density behind the shock. All measurements were non-dimensionalized using this induced velocity ($u_{shock}$) and density ($\rho_{shock}$) behind the shock and the characteristic length (*d*) which is considered to be the electrode gap distance (5 mm). The velocity and length scales were then used to define a characteristic time scale ($\tau$) as the electrode gap distance (*d*) divided by the velocity induced behind the shock. For the maximum energy value of 7 mJ, $u_{shock}$, $\rho_{shock}$ and $\tau$ are 71 m/s, 1.45 kg/m³ and 0.07 ms, respectively, and for the minimum energy value of 4 mJ, they are 54 m/s, 1.42 kg/m³ and 0.09 ms respectively.

## 3. Results and Discussion

### 3.1 Observations from experimental results

An example of the flow field induced after a single spark discharge generated under quiescent conditions, in a 5 mm electrode gap and corresponding to deposited energy of 5.8 mJ is presented in Figure 5. The flow is initially dominated by the formation of a thin region of hot gas in the gap between the two electrodes, the formation of a shock wave and its subsequent radial expansion, all within less than 30 µs ($<t/\tau = 1$)[50]. The flow field continues to evolve, and the measurements reported in this study are collected approximately 100 µs ($t/\tau > 1$) after the discharge, by which time the shock has departed the field of view. At $t/\tau = 1.9$, a cylindrical shaped gas kernel exists within the electrode gap as seen in the density measurements shown in Figure 5 and the density of the gas kernel is lower than the ambient, with $\rho/\rho_{shock} = 0.3$ (~ 40% of the ambient density). This hot gas kernel expands while simultaneously cooling over time, changing in shape from cylindrical, to the characteristic toroidal shape previously observed in visualization studies of flow induced by sparks [12], [15]. The hot gas kernel continues to cool over time until its density eventually equilibrates to the ambient.

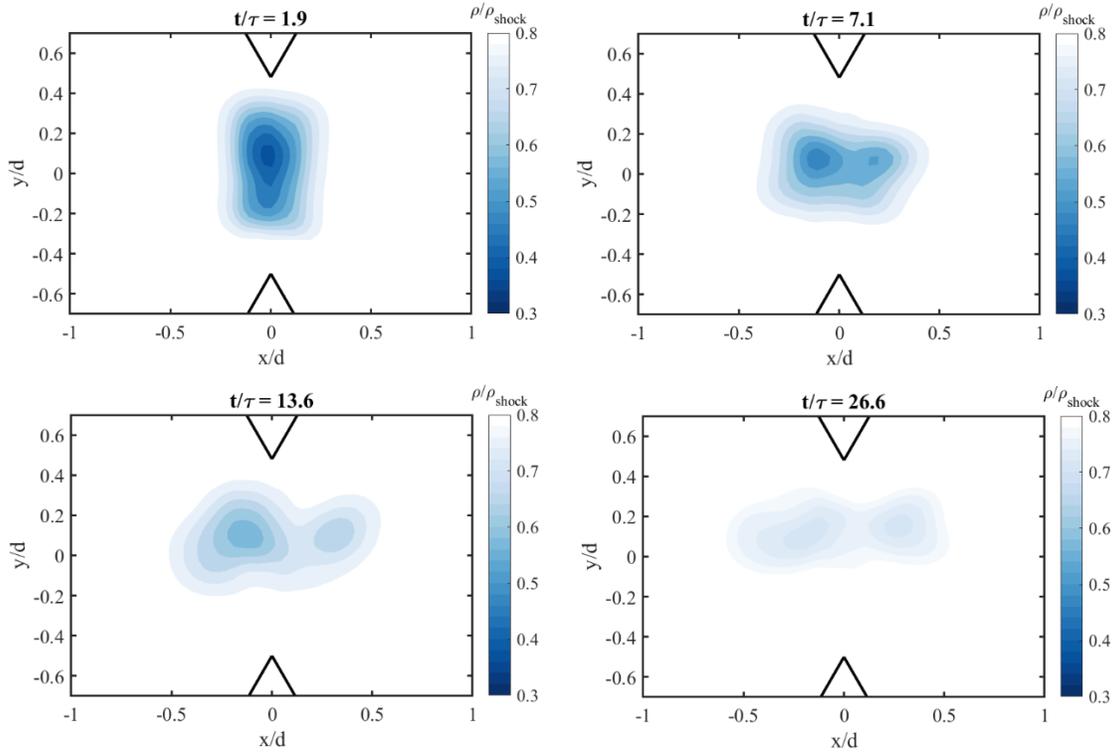

Figure 5: Time evolution of the density of the hot gas kernel from $t/\tau = 1.9$ (0.15 ms) to 26.6 (2.05 ms) obtained from BOS measurements. The measurements show minimum kernel densities ($\rho/\rho_{shock}$) as low as 0.3 and the evolution of the kernel shape from cylindrical to toroidal shape for a single test where $\rho_{shock} = 1.47 \ kg/m^3$ and $\tau = 0.08 \ ms$.

The mean density of all points lying within the hot gas kernel is computed, and a time history of the mean kernel density is shown in Figure 6(a). We observe an initial rapid increase in density from $\rho/\rho_{shock} = 0.64$ to 0.75, followed by a slower increase in density. The increase in density represents a cooling of the hot gas kernel, and the two stages of cooling both show linear increase in density. For the current test, we see that the rate of cooling in the first stage ($m_{fast}$) is nearly 5 times larger than in the second stage ($m_{slow}$), and the changeover ($t_{cp}$) from the first stage to the second occurs at $t/\tau = 8.8$.

This analysis is repeated for all 25 BOS experiments and the results are shown in Figure 6(b). We find that for each experiment two cooling regimes are present, a fast cooling regime and a slow cooling regime. Figure 6(b) also shows that for the spark energy ranges considered herein, the rates of cooling in the first (fast) regime are higher, by at least a factor of 2, than in the second, slow regime. In addition, neither the fast nor slow cooling rates display a clear dependence on the energy deposited in the gas by the discharge plasma.

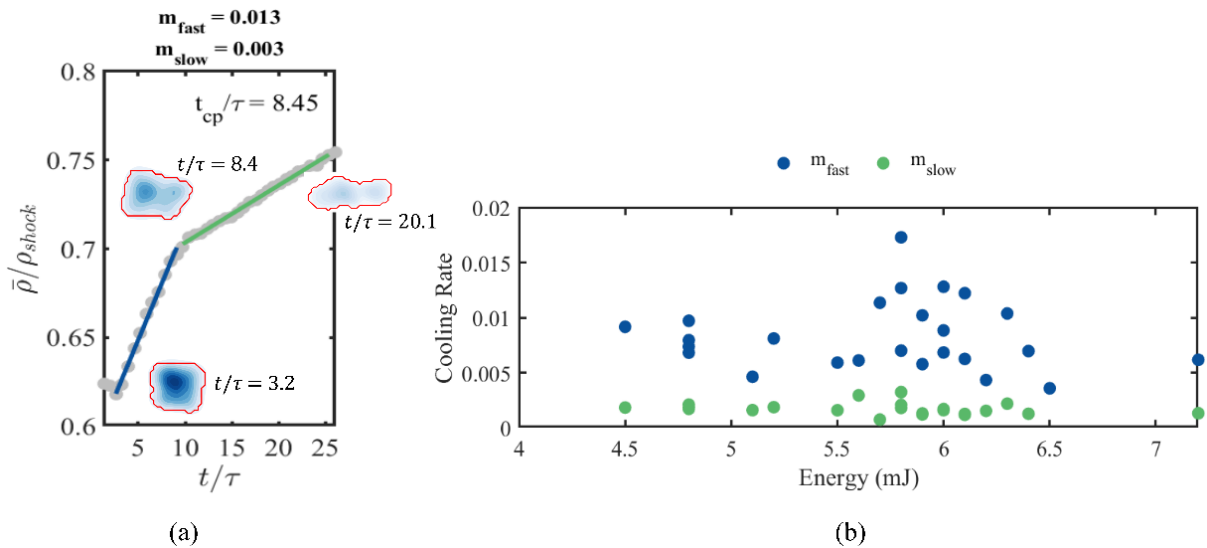

(a)  (b)

Figure 6: (a) Time history of mean kernel density inside BOS control volume, showing presence of two cooling regimes, each with different rates of cooling, with inset of identified kernel control volume from BOS data at $t/\tau$ values of 3.2, 8.4 and 20.1 (b) fast and slow cooling rates for all tests, showing no dependence on energy deposited.

Researchers have in the past hypothesized that the shape and density of the hot gas kernel is strongly coupled to the velocity field induced due to the spark [15]. Figure 7, shows an example of the vorticity field of the induced flow due to a single spark corresponding to deposited energy of 6.8 mJ. A pair of vortex rings are present near the electrode tips whose vortex cores are outlined by black circles and whose weighted centroids are represented by the filled circles contained inside the cores. Vortex trajectories of the pair of vortex rings in Figure 8(a) show that these migrate towards each other in the y-direction due to self-induction until they eventually collide [53]. As the rings get closer, the ring radii increase due to mutual induction as shown by the time history of the mean ring radii of the top and bottom rings in Figure 8(b). The mean circulation of both rings also continues to decay post collision as shown by the time history plot of circulation in Figure 8(b). Figure 9 shows the collision time of the vortex rings from S-PIV and the changeover time ($t_{cp}$) in the cooling rates measured from BOS. For the experiments considered, the median collision time in PIV is $t/\tau = 11.6$ and the median changeover time in BOS is $t/\tau = 10.4$.

As previously noted from the BOS data, all 25 tests showed that the hot gas kernel cools over time in two distinct cooling regimes. Similarly, all 25 S-PIV tests showed significant entrainment of ambient gas into the electrode gap, which may be responsible for the cooling observed in BOS. Based on these observations, we hypothesize that the entrained ambient fluid mixes with the hot gas kernel, and therefore the entrainment drives the cooling of the hot gas kernel. A model of the cooling process based on this hypothesis, is developed and tested in the following section.

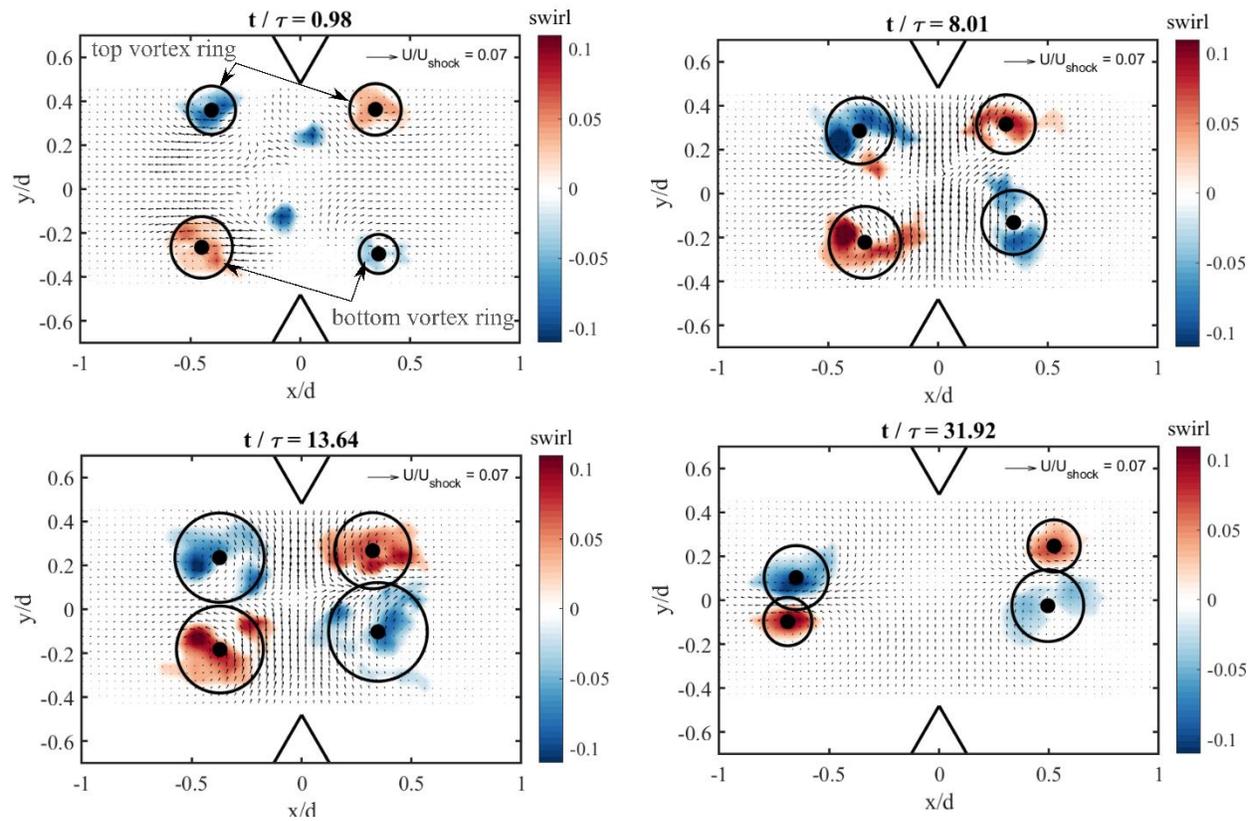

Figure 7: Coherent vorticity induced by the spark from t/$\tau$ = 0.98 (0.07 ms) to 27.7 (2.0 ms) obtained from PIV measurements. The results show motion of a pair of vortex rings towards each other and collision of the vortex rings followed by decay, for a single test where $u_{shock} = 66\ m/s$ and $\tau = 0.07\ ms$.

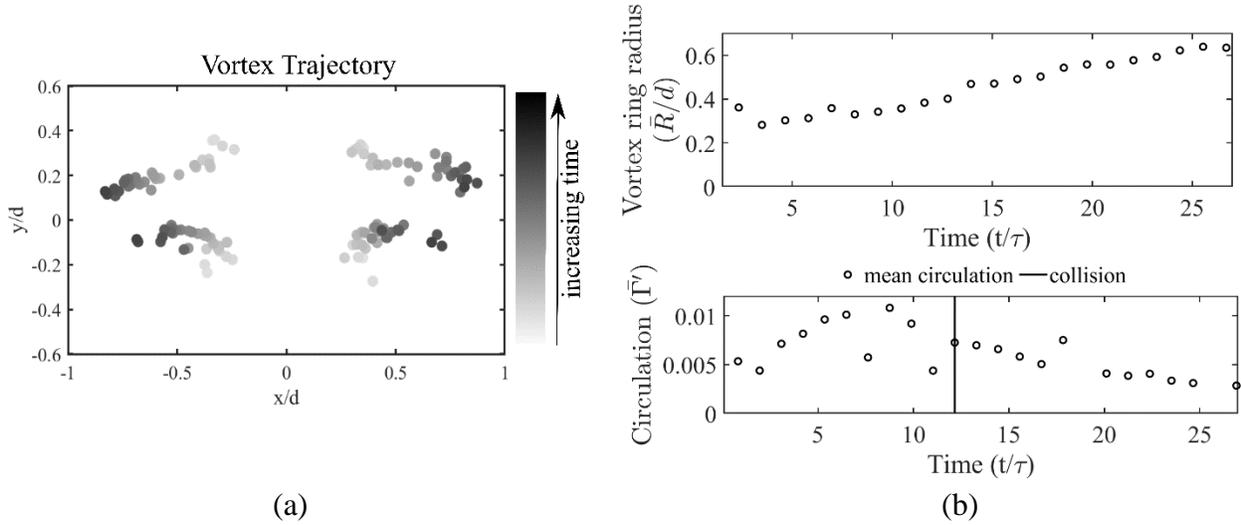

(a)                                        (b)

Figure 8: (a) Map of vortex trajectories showing motion towards y/d = 0 and simultaneous radial expansion in x. (b) Time history of mean radius of top and bottom vortex rings showing overall expansion and time history of mean circulation showing decay post collision of the vortex rings.

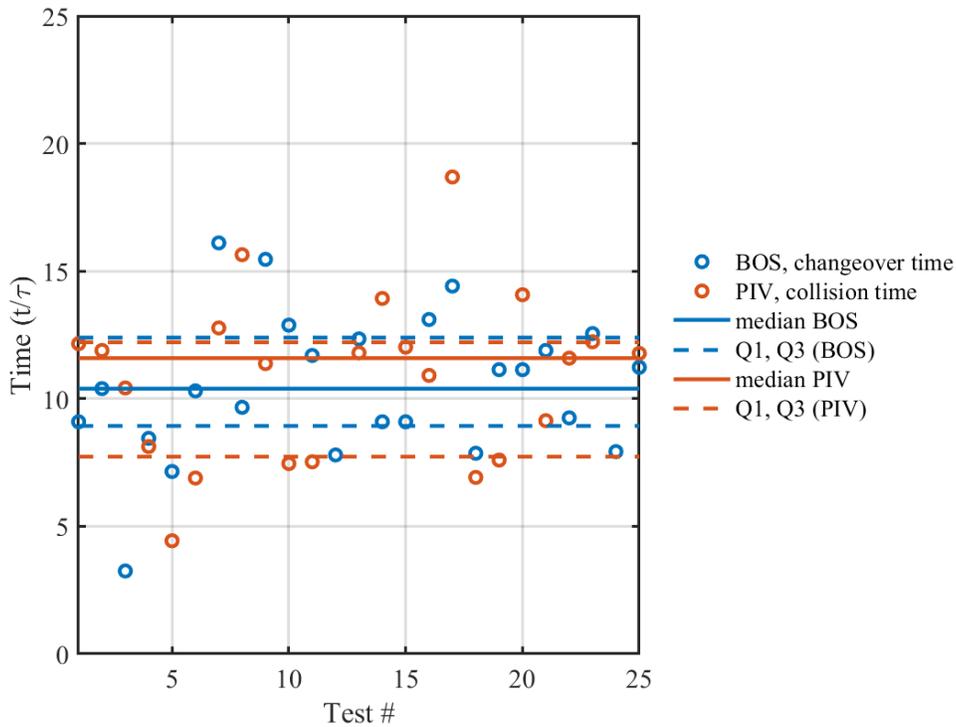

Figure 9: Comparison of cooling regime changeover points obtained from BOS with vortex ring collision times obtained from PIV. Quartiles of changeover points from the fast cooling to slow cooling range between $t/\tau = 8.9$ and 12.4 for BOS and quartiles of collision times range between $t/\tau = 7.7$ and 12.2 for PIV

### 3.2 A model for the effect of cold gas entrainment on the kernel cooling process

In order to test the hypothesis that entrainment of cold ambient gas is responsible for cooling of the hot gas kernel in the fast regime, a model is developed to relate the total cooling of the hot gas kernel to the volume of entrained fluid within a control volume representing the kernel. A pair of vortex rings is induced near each electrode tip and the PIV control volume is defined between the centroids of the cores of the vortex rings; details on this control volume definition will be provided in the following section. Cold ambient gas is entrained into the control volume, cooling the hot gas kernel.

The cooling process can be characterized by considering the rate of change of enthalpy of the kernel, with the simplifying assumption that the fluid is inviscid, thereby neglecting work done by viscous forces. Further neglecting body forces and heat transfer due to conduction and radiation enables the simplification of the total energy budget to the following equation,

$$\frac{\partial}{\partial t}\int_V (\rho h_0)\, dV + \iint \rho h_0\, \vec{u}\cdot \vec{n}\, dS = 0 \tag{6}$$

where $h_0$ is the stagnation (or total) enthalpy in the control volume and given by $h_0 \equiv e + \frac{P}{\rho} + \frac{1}{2}\vec{u}\cdot\vec{u}$. Here $e$ is internal energy (per unit mass), $\frac{P}{\rho}$ is work done due to pressure forces and $\frac{1}{2}\vec{u}\cdot\vec{u}$ is the kinetic energy (per unit mass).

Assuming a calorically perfect gas, the energy equation can be further simplified to

$$\frac{\partial}{\partial t}(mc_p T) = \rho_{in} c_p T_{in} \dot{Q}_{in} - \rho_{out} c_p T_{out} \dot{Q}_{out} \tag{7}$$

where $m$ is the mass of the hot gas kernel, $c_p$ is the specific heat capacity at constant pressure, and $T$ is the temperature of the hot gas kernel. Further, $\rho_{in}$ and $T_{in}$ represent the density and temperature of ambient fluid entering the control volume and conversely $\rho_{out}$ and $T_{out}$ are the density and temperature of fluid exiting the control volume. Finally, $\dot{Q}_{in}$ and $\dot{Q}_{out}$ denote the entrainment (influx) and detrainment (efflux), respectively.

The energy equation can be simplified even further by employing additional assumptions about the flow field. Assuming the flow field to be incompressible (as the cooling process occurs long after the shock has departed the field of view) implies that $\dot{Q}_{in} = \dot{Q}_{out}$. On considering air to be a perfect gas ($P = \rho RT$), changes in temperature due to pressure in the control volume are negligible. The temperature of the entrained ambient fluid is considered to be constant and at the ambient temperature under the assumption of an infinite reservoir of cold gas ($T_{in} = T_\infty = const$) in relation to the size of the control volume. It is also assumed that the entrained fluid mixes perfectly with the gas inside the kernel, and therefore the temperature of the fluid leaving the control volume, $T_{out}$, is considered to be equal to the temperature of the kernel, $T$.

Under these assumptions, the energy equation can be further simplified to relate the rate of change of density of the kernel to the entrainment, and is given by,

$$\frac{1}{(\rho_\infty - \rho)}\frac{\partial \rho}{\partial t} = \frac{\dot{Q}_{in}}{V_{cv}} \tag{8}$$

where the left-hand side of the equation is obtained from BOS measurements and the right-hand side is obtained from the S-PIV measurements. Here $V_{cv}$ is the volume of the control volume defined between the centroids of the coherent structures, as described in the following section.

Thus, the change in the mean kernel density can be calculated by integrating each side of the equation,

$$\int_{\overline{\rho_i}}^{\overline{\rho_f}} \frac{d\rho}{\rho - \rho_\infty} = \int_{t_i}^{t_{collision}} \frac{\dot{Q}_{in}}{V_{cv}} dt \tag{9}$$

where the integral on the left represents the total change in kernel density (the cooling effect) observed in the fast regime of the experiment, integrated from the initial mean kernel density (minimum of time history of mean kernel density) to the density at the changeover point, and the right-hand side represents the total amount of cold gas that is entrained as a fraction of the volume of the hot gas kernel, integrated from time $t/\tau > 3$, corresponding on average to the time of minimum density from BOS, to the time of collision. A metric termed the *cooling ratio* is introduced at this point and is obtained by further subtracting both sides of the equation from 1, as given by Equation (10):

$$\left(\frac{\overline{\rho_f} - \overline{\rho_i}}{\rho_\infty - \overline{\rho_i}}\right) = 1 - exp\left(-\int_{t_i}^{t_{collision}} \frac{\dot{Q}_{in}}{V_{cv}} dt\right) \tag{10}$$

It can be observed from the equation that for a fixed initial density $\overline{\rho_i}$, as the final density of the kernel $\overline{\rho_f}$ approaches the ambient density $\rho_\infty$, the ratio on the left-hand side of the equation (the 'cooling ratio') approaches 1. Therefore, the cooling ratio represents the extent of completion of the cooling process, and the closer its value is to 1, the more complete the cooling process.

*3.3 Calculation of parameters in the cooling model*

As stated earlier, the two sides of Equation (10) will be estimated from separate measurements, with the left-hand side evaluated using the density measurements from BOS, and the right-hand side using velocity measurements from S-PIV. One of the limitations of this calculation procedure is that the measurements are not simultaneous, and since the spark discharge is a chaotic process which is very sensitive to initial conditions (such as the electrode tip geometry and surface roughness), each realization of the spark induced flow field will be different. Our objective in this study is to therefore perform a statistical analysis of the quantities on the two sides of the equation and assess their agreement.

To facilitate this comparison, we denote the cooling ratio on the left-hand side of Equation (10) as $\kappa_{BOS}$, and is defined by:

$$\kappa_{BOS} = \left(\frac{\bar{\rho}_f - \bar{\rho}_i}{\rho_\infty - \bar{\rho}_i}\right) \tag{11}$$

Similarly, the term on the right hand side of the equation is denoted by $\kappa_{PIV}$, and is defined by:

$$\kappa_{PIV} = 1 - exp\left(-\int_{t_i}^{t_{collision}} \frac{\dot{Q}_{in}}{V_{cv}} dt\right) \tag{13}$$

In order to calculate the BOS cooling ratio $\kappa_{BOS}$, the initial and final densities in the fast cooling regime are determined from the time histories of the mean kernel density, as shown in Figure 6(a-b). The initial density is considered to be the absolute minimum mean density measured during the time series, while the final density is considered to be the density at the changeover time in BOS, which marks the transition from the fast to the slow cooling regime.

The term $\kappa_{PIV}$ represents the total entrainment/volume flux into an arbitrary control volume ($V_{cv}$) representing the hot gas kernel, here defined to be a polygon with the vertices located on the swirl strength weighted centroids of the coherent structures identified. Entrainment is calculated over the time period before vortex ring collision which marks the changeover from fast to slow cooling regimes in the S-PIV data. The velocity of the fluid relative to the control volume is used to calculate the entrainment, where the velocity of an edge of the control volume is calculated by taking a finite difference of the time history of each point along the boundary of the control volume. The velocity of the fluid at the edge of the control volume is calculated by interpolating the velocity measurements from the S-PIV grid on to the control volume edge of interest. This procedure is repeated for all edges of the control volume, and any influx of fluid is classified as entrainment. The volume of the control volume is defined as $V_{cv} = \pi \frac{w^2}{4} h$, where $w$, the width of the control volume, is defined as the mean of the width of the top and bottom boundaries, and $h$, the height, is defined as the mean y distance of the left and right control volume boundaries.

As the model described in this work is data-driven, and all measurements contain inherent uncertainties, it is essential to propagate these uncertainties in the raw measurements (velocities and densities) through the calculation procedure associated with the cooling model. The uncertainties in the fundamental quantities such as the velocity and density are propagated through each quantity in the cooling ratio equation using the Taylor series based propagation method [54] to determine the final uncertainty in cooling ratios estimated from BOS and S-PIV. The uncertainties in cooling ratio from the BOS and S-PIV measurements ranged from 1% to 16% of the cooling ratio value, with the majority (~ 80%) of the cooling ratio uncertainties being less than 5% of their respective cooling ratio values. Details of the uncertainty propagation are provided in the appendix.

### 3.4 Results of the data driven cooling model

The cooling ratios calculated from both S-PIV and BOS for all tests are shown in Figure 10. While there is considerable scatter in the estimates from both measurements, it is seen that the S-PIV cooling ratios $\kappa_{PIV}$ are distributed about a median value of 0.41, while the BOS cooling ratios $\kappa_{BOS}$ exhibit a median value of 0.39. The 25$^{th}$ and 75$^{th}$ quartiles show that these cooling ratio values range from 0.32 to 0.44 for BOS and from 0.3 to 0.51 for S-PIV. This implies that approximately 50% of the cooling occurs in the fast regime which spans about a millisecond, and illustrates the rapid, convective nature of the cooling process. As seen from Figure 10, the cooling ratios separately calculated from BOS and S-PIV are statistically equivalent over the range of experiments considered in this work, thus supporting the hypothesis that the cooling of the kernel is governed by the entrainment of cold gas.

It is important to note, however, that diffusive effects will also play a role in the kernel cooling in the fast regime, albeit at a much slower rate. The scatter in the cooling ratio estimates arise from the variability of each experiment, due primarily to the stochastic nature of the spark. This therefore prevents an exact match of cooling ratios from the two experimental campaigns.

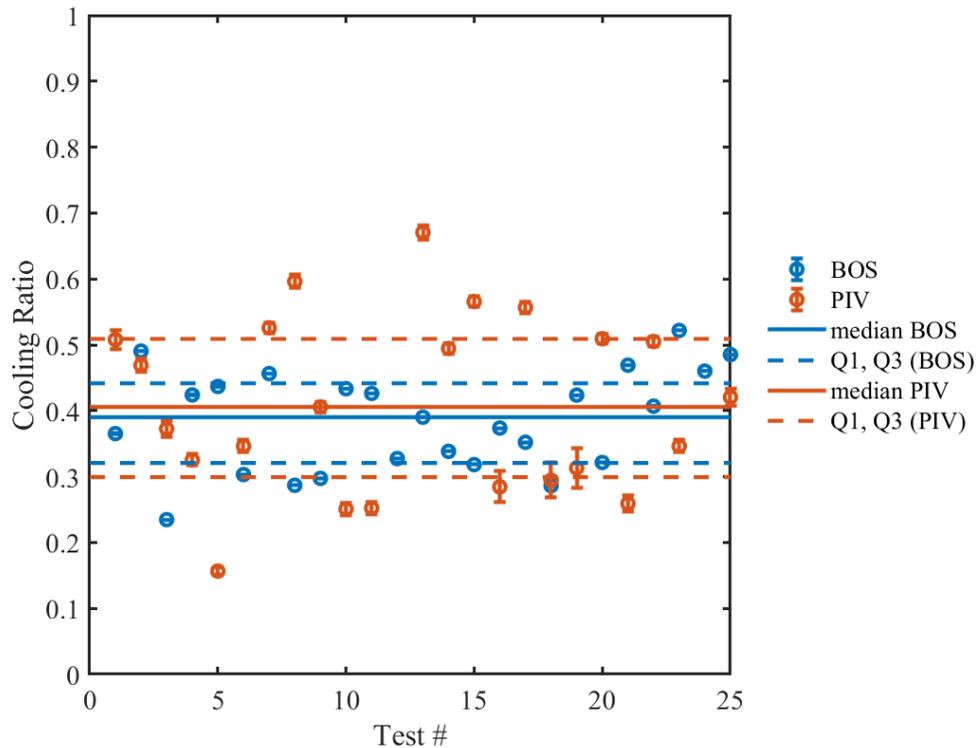

Figure 10: Cooling ratios calculated from S-PIV and BOS measurements show that in the fast regime, the total cooling effect observed from the kernel density measurement is statistically equivalent to the ratio measured from the cold gas entrained.

## 4. Conclusions

This work uses stereoscopic particle image velocimetry (S-PIV) and background oriented schlieren (BOS) measurements to characterize the flow induced by a spark generated between two conical shaped electrodes. The density field measured from BOS shows the cooling and expansion of an induced hot gas kernel, transforming in shape from an initially cylindrical region of hot gas to a torus shape, while simultaneously cooling to the ambient temperature/density. It was also observed that the cooling process, characterized by an increase in the mean kernel density, occurs in two stages: an initial fast, convective cooling regime on the order of a millisecond, followed by a slower cooling regime. Additionally, it was found that the rate of increase of density in the fast regime was at least twice as high as that in the slow regime for all 25 tests. The velocity measurements from S-PIV show the presence of a pair of vortex rings near each electrode tip and significant entrainment of fluid into the electrode gap. The vortex rings move towards each other and eventually collide, followed by the subsequent decay of vorticity.

Based on the observations, it is hypothesized that the entrainment of cold ambient gas drives the cooling of the hot gas kernel. In order to test this hypothesis, a simplified model of the cooling process was developed by considering the conservation of thermal energy inside a control volume representing the hot gas kernel. It is shown from the model that the rate of increase in density is directly proportional to the cold gas entrainment per unit volume of the kernel, and a metric termed the 'cooling ratio' was proposed to characterize the extent of this cooling process. From the point of view of the kernel density, the cooling ratio represents the ratio of the change in kernel density over the fast regime to the initial density deficit with respect to the ambient. From the point of view of entrainment, the cooling ratio represents the time integral of the entrainment of cold gas per unit volume of the hot gas kernel, over the duration of the fast regime.

The cooling model was tested by calculating the cooling ratio terms separately from the BOS and S-PIV measurements conducted over different experimental campaigns. The ratios were seen to be statistically equivalent within the limits imposed by the measurement uncertainty and the chaotic nature of the spark discharge, thereby confirming the hypothesis. The cooling ratios were distributed around median values of 0.39 and 0.41 for BOS and S-PIV respectively. This also implies that close to 50% of the cooling occurs in the fast regime which lasts just over a millisecond. This information can be used to inform the use of multiple pulses and pulse repetitive frequencies which aim to benefit from the synergetic effect of multiple plasma discharges [21], by ensuring that the frequency of these is always greater than $1/t_{cp}$. For the range of energies evaluated in these experiments, the frequencies needed to obtain synergetic effects between pulses have to be greater than 1 kHz. The cooling of the hot gas kernel also gives insight into the mixing process with the density representing passive scalar mixing. An initial driving velocity field during the entrainment process brings about significant mixing in the fast regime, resulting in cooling of the hot gas kernel. This would be beneficial when spark plasma discharges are used in combustible mixtures to improve combustion efficiency as well as reduce ignition delay.

Further improvements can be made to this analysis by conducting simultaneous PIV and BOS measurements so as to eliminate discrepancies in cooling ratio values for a single test due to the stochastic nature of the spark. A detailed analysis of the cause of entrainment that leads to this cooling will also provide further insight into the flow induced and the cooling of the hot gas kernel. This understanding of the flow field and the relationship between fluid entrainment and the cooling process will inform future studies on these types of spark discharges and can be extended to understanding a broad range of phenomena related to all types of plasma discharges and their applications.

# Appendix

The uncertainty in the BOS cooling ratio $\kappa_{BOS}$ is calculated as,

$$\sigma_{\kappa_{BOS}} = \sqrt{\left(\frac{1-\bar{\rho}_i}{\rho_\infty - \bar{\rho}_i}\right)^2 \sigma_{\bar{\rho}_f}^2 + \left(\frac{\bar{\rho}_f - \rho_\infty}{(\rho_\infty - \bar{\rho}_i)^2}\right)^2 \sigma_{\bar{\rho}_i}^2} \qquad (A.1)$$

where, $\sigma_{\bar{\rho}}$ represents the uncertainty in the mean kernel density and can be expressed in terms of the density uncertainties of the points within the kernel as $\sigma_{\bar{\rho}} = \frac{1}{n}\sqrt{\sum \sigma_\rho^2}$. The uncertainties in the cooling ratio estimates from BOS measurements ranged from 0.04% to 0.09% of the cooling ratio value.

The uncertainty in the cooling ratio calculated using S-PIV data is similarly determined by propagating uncertainties through the right-hand side of the cooling ratio equation, and is given by,

$$\sigma_\kappa = \sqrt{\left(\frac{\partial \kappa}{\partial \eta}\right)^2 \sigma_\eta^2} \qquad (A.2)$$

where $\eta = \frac{\dot{Q}_{in}}{\frac{w^2}{4} h}$ is the instantaneous entrainment per unit volume of the kernel. The uncertainty in this term is given by,

$$\sigma_\eta = \sqrt{\left(\frac{\partial \eta}{\partial \dot{Q}_{in}}\right)^2 \sigma_{\dot{Q}_{in}}^2 + \left(\frac{\partial \eta}{\partial w}\right)^2 \sigma_w^2 + \left(\frac{\partial \eta}{\partial h}\right)^2 \sigma_h^2} \qquad (A.3)$$

where the entrainment is defined as,

$$\dot{Q}_{in} = \int_0^{\frac{w}{2}} \int_0^{2\pi} v_{rel} x \, dx \, d\theta \qquad (A.4)$$

the equation for relative velocity ($v_{rel}$) is given by the equation below,

$$v_{rel} = \sqrt{\left(v - v_{ring}\right)^2 + \left(u - u_{ring}\right)^2} \qquad (A.5)$$

giving the uncertainty in entrainment as,

$$\sigma_{\dot{Q}_{in}} = 2\pi \sqrt{\left(\frac{\partial \dot{Q}_{in}}{\partial v_{rel}}\right)^2 \sigma_{v_{rel}}^2 + \left(\frac{\partial \dot{Q}_{in}}{\partial x}\right) \sigma_x^2} \qquad (A.6)$$

The uncertainty in the relative velocity, calculated using a 4[th] compact noise optimized Richardson extrapolation finite difference scheme used in entrainment calculation is given by,

$$\sigma_{v_{rel}} = \sqrt{\left(\frac{\partial v_{rel}}{\partial v}\right)^2 \sigma_v^2 + \left(\frac{\partial v_{rel}}{\partial v_{ring}}\right)^2 \sigma_{v_{ring}}^2 + \left(\frac{\partial v_{rel}}{\partial u}\right)^2 \sigma_u^2 + \left(\frac{\partial v_{rel}}{\partial u_{ring}}\right)^2 \sigma_{u_{ring}}^2} \qquad (A.7)$$

with the uncertainties in the calculation of ring velocities calculated according to [48].

The uncertainties in calculation of the width and height of the control volume are dependent on the uncertainty in the centroid ($X_0$ and $Y_0$), given below,

$$\sigma_w = \frac{1}{\sqrt{2}} \sigma_{X_0} \qquad (A.8)$$

$$\sigma_h = \sqrt{2} \sigma_{Y_0} \qquad (A.9)$$

$$\sigma_{X_0} = \sqrt{\left(\frac{1}{\sum \lambda_{ci_i}}\right)^2 \left(\sum \sigma_{\lambda_{ci_i}}^2 x_i^2 + \sigma_{x_i}^2 \lambda_{ci_i}^2\right) + \left(\frac{\sum x_i \lambda_{ci_i}}{\left(\sum \lambda_{ci_i}\right)^2}\right)^2 \left(\sum \sigma_{\lambda_{ci_i}}^2\right)} \qquad (A.10)$$

$$\sigma_{Y_0} = \sqrt{\left(\frac{1}{\sum \lambda_{ci_i}}\right)^2 \left(\sum \sigma_{\lambda_{ci_i}}^2 y_i^2 + \sigma_{y_i}^2 \lambda_{ci_i}^2\right) + \left(\frac{\sum y_i \lambda_{ci_i}}{\left(\sum \lambda_{ci_i}\right)^2}\right)^2 \left(\sum \sigma_{\lambda_{ci_i}}^2\right)} \qquad (A.11)$$

where $\lambda_{ci}$ is calculated as,

$$\lambda_{ci} = \left| imag\left(\frac{1}{2}\left(\frac{\partial u}{\partial x} + \frac{\partial v}{\partial y}\right) \pm \frac{1}{2}\sqrt{\left(\frac{\partial u}{\partial x} + \frac{\partial v}{\partial y}\right)^2 - 4\left(\frac{\partial u}{\partial x}\frac{\partial v}{\partial y} - \frac{\partial u}{\partial y}\frac{\partial v}{\partial x}\right)}\right) \right| \qquad (A.12)$$

and its uncertainty ($\sigma_{\lambda_{ci}}$) is given by,

$$\sqrt{\left(\frac{1}{4}\left(\frac{0.7085}{\Delta}\right)^2 (\sigma_u^2 + \sigma_v^2) + \left(\frac{1}{\left(\sqrt{\frac{1}{4}\left(\frac{\partial u}{\partial x}\right)^2 + \frac{1}{4}\left(\frac{\partial v}{\partial y}\right)^2 - \frac{1}{2}\frac{\partial u}{\partial x}\frac{\partial v}{\partial y} + \frac{\partial u}{\partial y}\frac{\partial v}{\partial x}}\right)^2}\left(\frac{1}{4}\left(\frac{\partial u}{\partial x}\right)^2 + \frac{1}{4}\left(\frac{\partial v}{\partial y}\right)^2 + \left(\frac{\partial v}{\partial x}\right)^2\right)\left(\frac{0.7085}{\Delta}\right)^2 \sigma_u^2 + \left(\frac{1}{4}\left(\frac{\partial v}{\partial y}\right)^2 + \frac{1}{4}\left(\frac{\partial u}{\partial x}\right)^2 + \left(\frac{\partial u}{\partial y}\right)^2\right)\left(\frac{0.7085}{\Delta}\right)^2 \sigma_v^2\right)\right)} \qquad (A.13)$$